# A first look at galaxies with the DENIS survey

GARY MAMON[1,2], VINCENT BANCHET[1], CATHERINE BOISSON[2], VÉRONIQUE CAYATTE[2], FRÉDÉRIC ENGELMANN[1]

[1] *Institut d'Astrophysique de Paris, Paris, France*
[2] *DAEC, Observatoire de Paris, Meudon, France*

**ABSTRACT.** We review the various steps for the extraction of extended sources, principally galaxies, from the DENIS images. The capabilities of DENIS are studied with simulated images. A few real DENIS $J$ and $K_s$-band images are presented for very bright galaxies and one nearby rich galaxy cluster.

## 1. Introduction

The DENIS Near-Infrared imaging survey of the southern hemisphere has important applications for extragalactic astronomy and cosmology, stemming in part from the advantage of near-IR (hereafter, NIR) light 1) to suffer little extinction by interstellar dust; and 2) to be better correlated with the stellar mass content of galaxies than bluer light (more enhanced by recent star formation in the galaxies). Besides providing the community with the first very large digital multi-waveband survey of a significant fraction of the sky, the galaxy lists that will be extracted from the DENIS images will allow 1) the statistical study of near-IR (hereafter, NIR) properties of galaxies; 2) the study of the distribution of galaxies throughout most of the southern hemisphere, that is down to very low galactic latitudes, and in some cases, right through the Galactic Plane (Mamon 1994); 3) the systematic study of bulge/disk ratios and their correlation with the environment (through $I - J$ colors obtained directly from DENIS, and $B - J$ colors obtained by comparison with optical surveys). The color segregation that should be found will be a good tracer of the variations of the bulge to disk ratio of galaxies with the environment, and the errors of the method should be offset by the formidable statistics that we expect ($> 10^5$ galaxies).

## 2. DENIS capabilities

2.1. General strategy

We simulate images that reproduce the basic characteristics of the DENIS images (but are perfectly flat-fielded, with no glitches, ghosts, etc.).

We can then estimate our *selection functions* for galaxy extraction, as a function of observing conditions (waveband, background dispersion and point-spread-function), field-of-view (confusion by stars and interstellar extinction), object parameters (magnitude, type and inclination). We then optimize our algorithmic parameters in terms of those external parameters.

This is a difficult task because of the large number of parameters, both external (basically 8, see above) and algorithmic (basically a smoothing scale for detection, a flag to use smoothed images or not for star/galaxy separation, a detection threshold, a star/galaxy separation threshold and aperture, a photometric aperture or threshold or both). We thus have seeked short cuts. First (as suggested to us by D. Lynden-Bell), we focus at high galactic latitudes, and we later intend to mimic the effects of confusion by adding a large number of artificial stars with the same PSF to our high-galactic latitude images, while the effects of increased extinction will be handled by increasing the background noise and/or decreasing the flux around the objects detected at high $b$. The goal would be to attempt to recover the high-$B$ galaxies.

## 2.2. Detection

Our detection algorithm consists of smoothing the image, thresholding and requiring a minimum number of connected pixels. This last quantity is set by requiring that the number of false detections in simulated images with no objects is less than some threshold. Note that this threshold of the frequency of false detections must be much more severe for galaxies than for stars, since the former are much rarer.

We have noticed that the optimal detection parameters depend little on the instrumental parameters, and on the object type, with the sole exception that face-on late-type (negligible bulge) spiral galaxies require a larger smoothing length in the $K_s$ band. In particular there is little difference between the optimal boxcar smoothing parameters and the optimal gaussian smoothing ones (more complicated smoothing filters do not enhance our detection, see Banchet *et al.* 1995). Moreover, there is little difference between 4-point and 8-point connectivity.

The magnitude limits for 50% and/or 95% completeness in detection depend little on the PSF, but vary with the background level, as $m_{\text{lim}} = \text{Cst} + \eta \mu_{\text{bg}}$, where $\mu_{\text{bg}}$ is the surface magnitude of the background. The slope $\eta$ is generally found to be near 0.5 as expected for point sources, and is closer to unity for late-type face-on spirals in the $K_s$ band.

## 2.3. Photometry

At least two types of photometry will be required on DENIS galaxies: fixed circular aperture photometry for color information, and an optimal photometry. Isophotal and iterative elliptical aperture methods are currently being tested with simulated images. For isophotal photometry, mean offsets and dispersions are found to be a strong function of galaxy type.

## 2.4. Star/Galaxy Separation

There are two broad classes of star/galaxy separation algorithms: 1) classical algorithms which rely on one or more well-defined statistics to plot clouds of points in 2 or more dimensions (one of which is usually the magnitude of the object), and a well-defined rule to separate this cloud of points into 2 classes; 2) neural networks, which after a period of training are meant to optimize the classification with hidden statistics and separation rules. We have chosen to follow both procedures, the latter one being supervised by E. Bertin. Because a set of images of roughly constant star-density within the same observed strip will contain too few stars in each magnitude bin, we have decided to define the stellar locus on the simulated images (for given observing band, PSF, and sky brightness). We can then define a distance to this stellar locus, in units of the standard deviation of this stellar locus, and later quantify the reliability of the galaxy classification in terms of this distance and the observed overall star-density.

## 3. Two bright galaxies

We have searched among roughly $100\,\mathrm{deg}^2$ of flat-fielded DENIS slots (available on disk at that time) for those with the brightest galaxies in the RC3 (de Vaucouleurs *et al.* 1991) galaxy catalog. In Figure 1, we present two among the brightest galaxies from our search: NGC 1512 (Sa, $B_T = 11.1$) and NGC 1515 (Sbc, $B_T = 11.7$). The $J$ band image of NGC 1512 shows the bar reaching to the ring, while the $K_s$ image of the galaxy shows traces of the bar and hints of the ring to the Northeast. In the $J$ image of NGC 1515, one can see the spiral arms, while only traces of them can be seen in the corresponding $K_s$ image. Note also that for this edge-on spiral, the $J$ image almost reaches out to the $\mu_B = 25$ isophote.

## 4. A nearby cluster

The two galaxies shown in Figure 1 are extreme examples. For more reliable pictures, we have also searched our $\simeq 100\,\mathrm{deg}^2$ of flat-fielded DENIS slots for the cluster with the brightest tenth galaxy magnitude, as given in the ACO (Abell, Corwin & Olowin 1989) catalog. In Figure 2, we show two fields of Abell 3223 (at redshift $z = 0.045$, and whose tenth brightest galaxy is of magnitude $R_{10} = 15.6$) for a qualitative assessment of the DENIS images. Note the two bright galaxies in the southern fields. We are in the midst of making a detailed comparison with COSMOS ($B$-band) lists of objects in these two fields (under the supervision of C. Boisson and V. Cayatte). First indications are that most of the fuzzy objects in Figure 2 are galaxies.

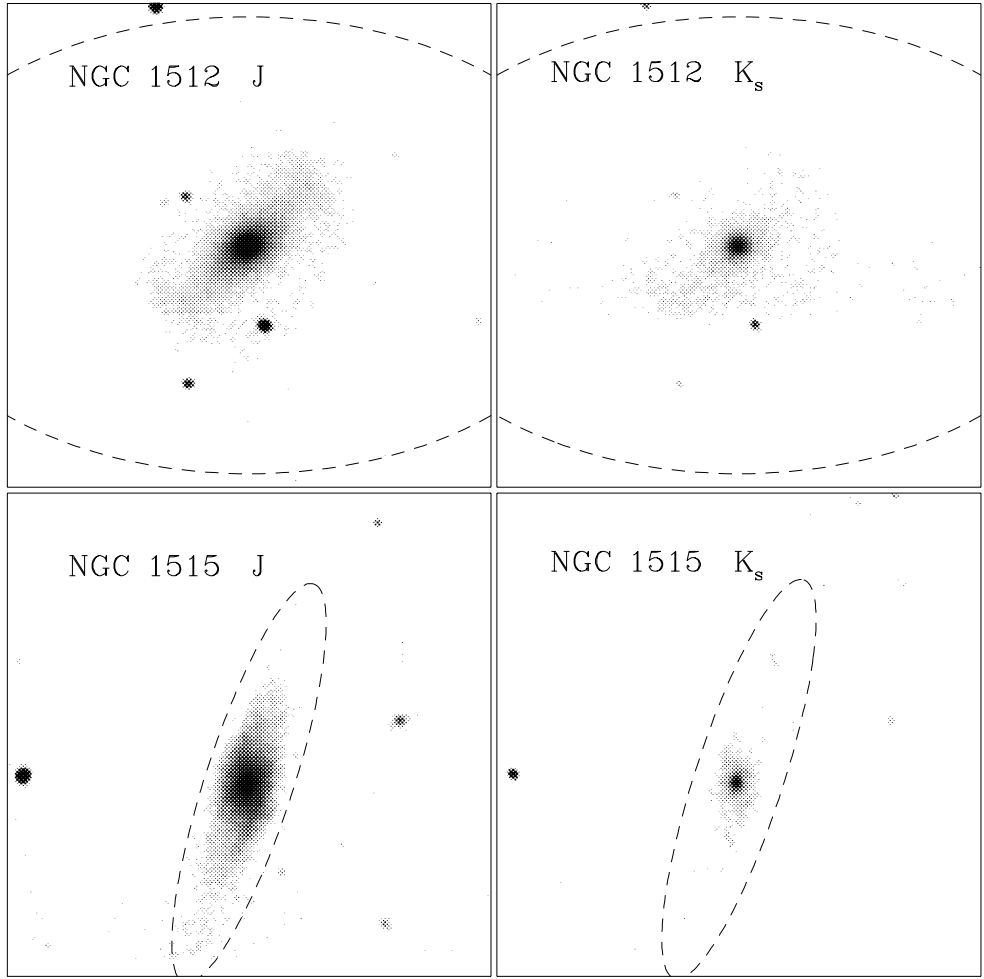

Fig. 1. $J$ (*left*) and $K_s$ (*right*) $3 \times 3$ boxcar smoothed images of NGC 1512 (*top*) and NGC 1515 (*bottom*). The fields are $6'$ wide. The ellipses are the 25th isophote in $B$. The lowest greyscale are $2\sigma$ for NGC 1512 and $3\sigma$ for NGC 1515 (where the dispersion $\sigma$ is of the smoothed image) and the greyscale increases with surface magnitude. North is at the bottom, East to the left.

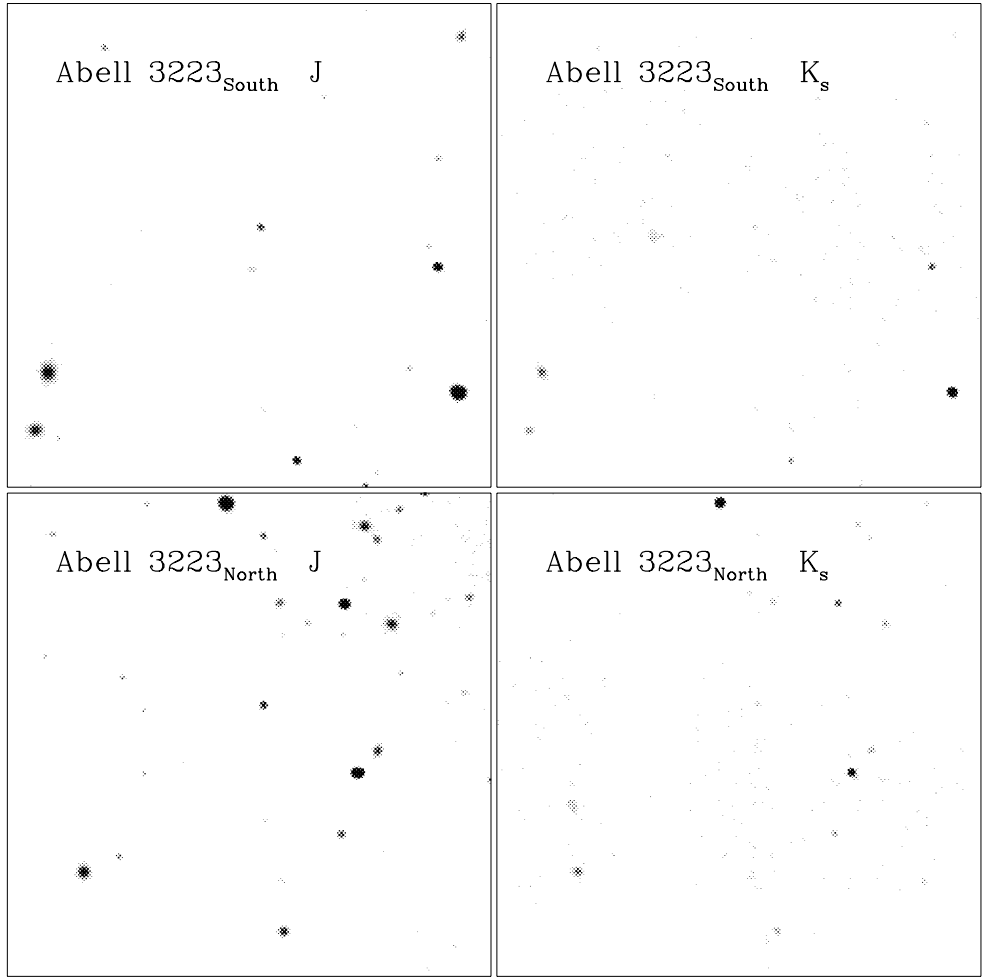

Fig. 2. $J$ (*left*) and $K_s$ (*right*) 3 × 3 boxcar smoothed images of Abell 3223. The fields are 6′ wide. The lowest greyscale is $3\sigma$ (where the dispersion $\sigma$ is of the smoothed image) and the greyscale increases with surface magnitude. North is at the bottom, East to the left. The northern and southern fields are a few arcmin apart in Dec.